\documentclass[twocolumn,english,aps,prl,noshowpacs,floatfix,superscriptaddress]{revtex4}
\usepackage[T1]{fontenc}
\usepackage[latin9]{inputenc}
\usepackage{amsmath}
\usepackage{amssymb}
\usepackage{graphicx}
\usepackage{float}

\makeatletter
\@ifundefined{textcolor}{}
{%
 \definecolor{BLACK}{gray}{0}
 \definecolor{WHITE}{gray}{1}
 \definecolor{RED}{rgb}{1,0,0}
 \definecolor{GREEN}{rgb}{0,1,0}
 \definecolor{BLUE}{rgb}{0,0,1}
 \definecolor{CYAN}{cmyk}{1,0,0,0}
 \definecolor{MAGENTA}{cmyk}{0,1,0,0}
 \definecolor{YELLOW}{cmyk}{0,0,1,0}
}

\newcommand*{\balancecolsandclearpage}{%
  \close@column@grid
  \clearpage
  \twocolumngrid
}

\@ifundefined{definecolor}
 {\usepackage{color}}{}
\usepackage{siunitx}

\makeatother

\usepackage{babel}
\begin{document}
\flushbottom

\title{Accuracy and minor embedding in subqubo decomposition with fully connected 
large problems: a case study about the number partitioning problem}% and qbsolv}

\author{Luca Asproni}

\affiliation{Data Reply s.r.l., Via Nizza, 262, 10126, Turin, ITALY}

\author{Davide~Caputo}

\affiliation{Data Reply s.r.l., Via Nizza, 262, 10126, Turin, ITALY}
\affiliation{Department of Mathematics and Physics - University of Salento, Via Arnesano, 73100, Lecce, ITALY}

\author{Blanca Silva}

\affiliation{Data Reply s.r.l., Via Nizza, 262, 10126, Turin, ITALY}

\author{Giovanni~Fazzi}

\affiliation{Data Reply s.r.l., Via Robert Koch, 1/4, 20152, Milan, ITALY}

\author{Marco~Magagnini}

\affiliation{QUANTUM COMPUTING COMMUNITY OF PRACTICE, Reply S.p.A., Corso Francia, 110, 10143, Turin, ITALY}

\begin{abstract}
\section*{Abstract}

In this work we investigate the capabilities of a hybrid quantum-classical procedure to explore the solution space using the D-Wave $2000Q^{TM}$ Quantum Annealer device.
Here we study the ability of the Quantum hardware to solve the Number Partitioning Problem, a well-known NP-Hard optimization model that poses some challenges typical of those encountered in real-world applications. 
This represents one of the most complex scenario in terms of qubits connectivity and, by increasing the input problem size, we analyse the scaling properties of the quantum-classical workflow.
We find remarkable results in most instances of the model; for the most complex ones, we investigate further the D-Wave Hybrid suite. Specifically, we were able to find the optimal solutions even in the worst cases by fine-tuning the parameters that schedule the annealing time and allowing a pause in the annealing cycle.
\end{abstract}

%\date{\today}
\maketitle

\section{introduction}

Recently, the availability for the first time of quantum annealing devices from 
D-Wave Systems has captured the attention of both researchers and 
technology companies 
\cite{booth2018comparing, venturelli2018compiling, amir2018, stollenwerk2017quantum}.
Besides, a growing interest is in the experimental determination
of whether or not a quantum speedup can be achieved with this new class
of quantum devices and what kind of working applications can be developed 
on such platforms 
\cite{PhysRevX.5.031040, 10.3389/fict.2017.00029,NIPS2017_7143, hamze2017}.

The participation of major technology powers such as Google, Lockheed Martin, 
and Los Alamos Laboratories continues rising, together with the scientific literature
and application reports.
Nevertheless, there is still a strong limitation in the usage of this model of computation 
for solving real world problems due to the limited number of qubits 
and couplers inside the quantum processor unit (QPU).
Actually, it is well known that quantum annealers need to have a dramatically larger number of qubits and couplers in order to model the complexity of real life problems.
Especially, the limited connectivity between qubits inside the current Chimera graph architecture
represents an additional obstacle in mapping large real problems in the QPU 
\cite{Chancellor2016, Chancellor2016_2, Pudenz2014, ojas2016}.

Furthermore, with the release of an open-source suite spanning from the decomposing solver Qbsolv to the new Hybrid framework, D-Wave took a significant step forward towards gathering the attention from technology companies. As a matter of fact, with these technologies it is possible to close the gap between logical qubits representation encoded in the QUBO (Quadratic Unconstrained 
Binary Optimization) matrix and the physical embedding of the problem into the Chimera 
graph \cite{Kumar2018}.

Besides, it is possible to decompose large problems into smaller subsets in such a way that they can be integrated immediately into the QPU, by providing both the combinatorial implementation required for the physical embedding and the decomposition procedure for the creation of the smaller instances. Also, the backend to be used during the computation can be specified in order to solve the model by means of either a classical or a quantum-based platform.

However, despite all the attention drawn to this crucial tool, a systematic investigation of the efficiency related to the optimization and decomposition performance has not been 
exhaustively conducted yet. Some studies have been developed using special techniques such as the time-to-target metric \cite{king2015} or applying methods based on the matrix factorization \cite{malley2017} but without taking into account the capabilities of scaling up when the 
input size grows.

In this work we investigate the accuracy and the capability of the D-Wave 2000Q quantum annealer to solve problems with a significantly large input. To perform this study, we use one well-known NP-Hard model: the number partitioning problem (NPP) \cite{mertens2006easiest}. Thanks to the simplicity of this problem, it is easy to generate artificial problems of any size for which the optimal solution is known. Consequently, the measurement of the quality of the solution provided by the quantum annealer, along with the classical implementation of the tabu-search algorithm for the problem decomposition, will be possible even for large datasets.

\section{number partitioning problem and quantum annealing}

The number partitioning problem (NPP) is defined as the task of discriminating if 
a given set $S$ of positive integer numbers can be divided (partitioned) 
into two subsets $S_1$ and $S_2$ where the total sum of the elements in $S_1$ equals the total sum of the elements in $S_2$. Although the NPP is an NP-complete problem, the optimization version is considered NP-Hard and can be formulated in the following way: given a list of $N$ positive integers 
$\{a_1,a_2,...,a_N\}$, the solution consists in finding
a subset $A \subset \{a_1, a_2, ..., a_N\}$ such that the difference:

\begin{equation}
  D(A) = \bigg| \sum_{i \in A} a_i - \sum_{i \not \in A} a_i \bigg|,
\end{equation}
is minimized. Throughout this work, we will refer to this difference as the $\textit{delta}$ between the two subsets $A$ and $S\setminus A$. This problem is of both practical and theoretical importance: possible real applications span from multiprocessor pipeline scheduling
\cite{doi:10.1137/1035026}, where balancing and partitioning different resources can be crucial, to cryptography \cite{10.1007/3-540-49264-X_3}  
and all those problems requiring load balancing for I/O capacities, e.g. during 
databases processing \cite{Lewis:2008:NMS:1287844.1288008}.

The D-Wave device implements a quantum annealing heuristics to solve sampling, optimization and machine learning problems. Specifically, given a physical system composed of qubits, it is possible to define its Hamiltonian and initialize it in such a way that the lowest-energy state corresponds to all qubits being in a superposition state of 0 and 1. Then, as the annealing proceeds, a new Hamiltonian deriving from the problem's specifications, called the problem Hamiltonian, is introduced and gradually takes over the initial energy landscape, up to a point where it contains all the energy contributions. The Hamiltonian of the system can be defined in the following way:

\begin{equation}
H(t) = H_I(t)+H_P(t),
\label{eq:0}
\end{equation}
where $H_I$ is the initial Hamiltonian, $H_P$ is the problem Hamiltonian and their temporal evolution through the annealing is such that $H_I(0)\gg H_P(0)$ and $H_I(t_f)\ll H_P(t_f)$, being $t_f$ the final time of the annealing. 

As the problem Hamiltonian is introduced, the energy levels of the excited states are originated, increasing the probability of the system to jump from the ground state to some other excited state. In particular, there exists a critical point, the point of minimum gap, where the ground-state energy level is closest to the lowest energy level of one of the excited states. In such point, the probability of escaping the ground state is the highest, in which case the system is driven away from the global minimum.

In practice, in order to manipulate the Hamiltonian of the system, an external magnetic field is applied to the qubits. In this way, the probability of qubits falling into the 0 or 1 state is changed. The quantity that controls the magnetic field, called bias or weight, is directly controlled by the function of the problem at hand, that is the one from which a sample is needed or that has to be minimized. Moreover, it is possible to correlate qubits by entangling them. This is obtained by setting the value of a coupler, which represents the strength of the correlation between qubits that are linked together.

Hence, by letting the initial system undergo the quantum annealing process, it is possible to raise energy barriers in such a way that the energy of the system reflects the function to be minimized or sampled from. If the quantum annealing is slow enough, the system is able to naturally end up in the lowest-energy state, i.e. the low energy states needed in a sampling problem or the solution of a minimization problem.

In its current implementation, the D-Wave's quantum annealer is able to solve problems expressed in the form of an Ising glass, with a Hamiltonian written in the following form: 

\begin{equation}
H = \displaystyle{\sum_{i=1}^N h_i S_i + \sum_{i \neq j} c_{ij}S_iS_j}
\label{eq:1}
\end{equation}
where $H$ is the Hamiltonian encoding the problem, $S_i \in \{-1, 1\}$ are the spin values and
$h_i$ and $c_{ij}$ are respectively the qubits weights and the couplers coefficients 
of the model.

A complete formulation of the NPP as Ising spin glass has been provided in 
Ref.\cite{10.3389/fphy.2014.00005}. The Hamiltonian for this type of problem can be defined by assuming an increase in the energy when the total of amplitudes associated with positive spin states is different from that of amplitudes with negative spins. 
Accordingly to this formulation, it is possible to use the following relation:

\begin{equation}
  H = \bigg( \sum_{i \in N} a_i S_i \bigg)^2
\end{equation}
with $S_i = \pm 1$ the spin values indicating the subset to which the i-\textit{th} 
element belongs and $a_i$ the element of the set $A$. 
It follows that if the ground state has $H>0$ there is no exact solution of the
specific problem and the ground state is the one minimizing the mismatch between the two 
subsets.

In order to formulate the problem as a Quadratic Unconstrained Binary Optimization model, we first have to convert our $S_i = \pm 1$ into binary variables of the form $q_i \in \{0,1\}$.
This can be done by using the following simple relation:  

\begin{equation}
  q_i = \displaystyle{\frac{S_i+1}{2}}
  \label{binary_transf}
\end{equation}
where $q_i$ is the $i$-th variable and $S_i$ is the spin value. Now, the original Ising problem can be mapped into the QUBO form:

\begin{equation}
  \text{min} \sum_{i,j}Q_{ij}x_ix_j
\end{equation}
where $x$ represents a binary variable and $Q$ is the so-called QUBO matrix containing the weights of qubits ($h_i$ in Eq.\ref{eq:1}) in the diagonal and the couplers coefficients ($c_{ij}$ in Eq.\ref{eq:1}) in the $(i,j)$ elements. This matrix will be symmetric ($c_ic_j=c_jc_i$), allowing a reduction in the number of variables by selecting only $i\leqslant j$ and setting the remaining terms to zero, leading to an upper-triangular matrix.

\begin{figure}%[htbp]
	\centering
	\includegraphics[width=0.48\textwidth]{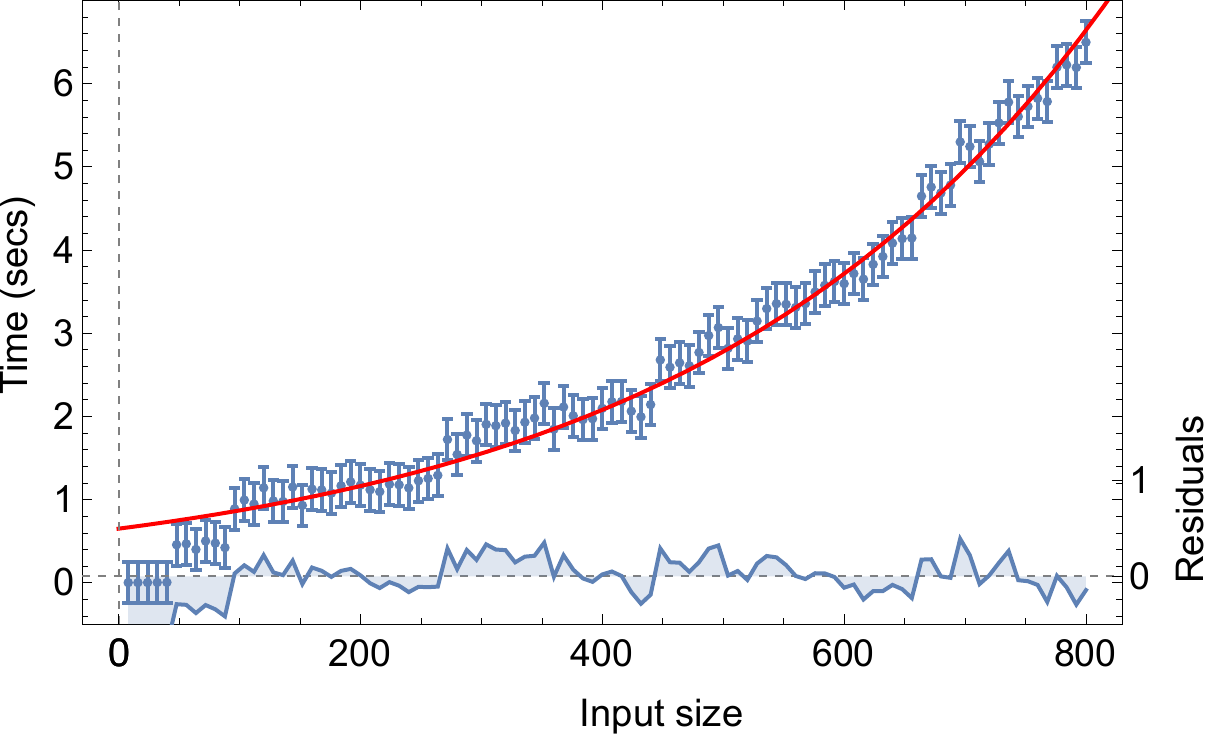}
	\caption{\textbf{Execution time of %Qbsolv 
			tabu-search for increasing input size}. 
		%Simulation using the number partitioning problem on classical hardware.
		Classical partitioning of a set with the classical embedded tabu-search 
		as backend. 
		The red line is the exponential fitting $t=A e^{x/B}$ where B=340 
		a.u. and x is the size of the input. Blue points represent measured data. 
		The blue line in the bottom part is the deviation of the experimental point from the
		value of the fitting.}
	\label{fig:1}
\end{figure}

Having the QUBO matrix, it is possible to submit it to the QPU and retrieve a solution of the optimization problem. However, the connectivity between qubits required by the NPP is that of a complete graph, which is yet to be supported by any modern quantum annealer that provides a fairly high number of qubits. To overcome this and similar problems, the D-Wave device operates a minor-embedding of the problem onto its Chimera architecture. Specifically, one can either run the built-in tabu-search heuristics provided by the D-Wave Hybrid tool to optimally decompose the problem into subproblems or choose a custom minor-embedding strategy. The subproblems will then be mapped onto che Chimera graph, for which the QPU will start the quantum annealing.

In Fig.~\ref{fig:1} the time required to solve the NPP on classical hardware by using the D-Wave Qbsolv is reported as a function of the input set size. The elaboration time increases exponentially while a structured procedure is applied in order to find the minimum: a number of subproblems are generated, handled and finally merged into a global solution of the NPP.
The exponential increase in the execution time confirms the NP-Hardness of the problem when approached with classical hardware and formulations.  When the problem is submitted to the QPU, the execution time changes and paves the way for a wide range of investigations of the D-Wave Hybrid tool. Moreover, this peculiar model allows us to study what happens in one of the worst case scenario from the perspective of the qubits connectivity: a fully connected graph, where the number of couplers and weights precision play a central role \cite{PhysRevX.6.031015, PhysRevX.5.031040}.

\section{results}\label{sec:exper}
In order to investigate the capabilities of the D-Wave hybrid tool, we solve multiple NPP examples of increasing size. For each fixed problem size we use 10 different datasets and collect statistics of the results. For experimental purposes, we choose the data in such a way that the ground state of the corresponding Ising models is $H=0$, i.e. there is a single partition of the set of numbers.

For our studies, we first construct the QUBO matrix for each problem, and then we define the tabu-search heuristics as the algorithm that splits the original problem into the subproblems, preparing them to be embedded on the Chimera graph.

\begin{figure*}%[htbp]
	\centering
	\includegraphics[width=0.3\textwidth, angle=270]{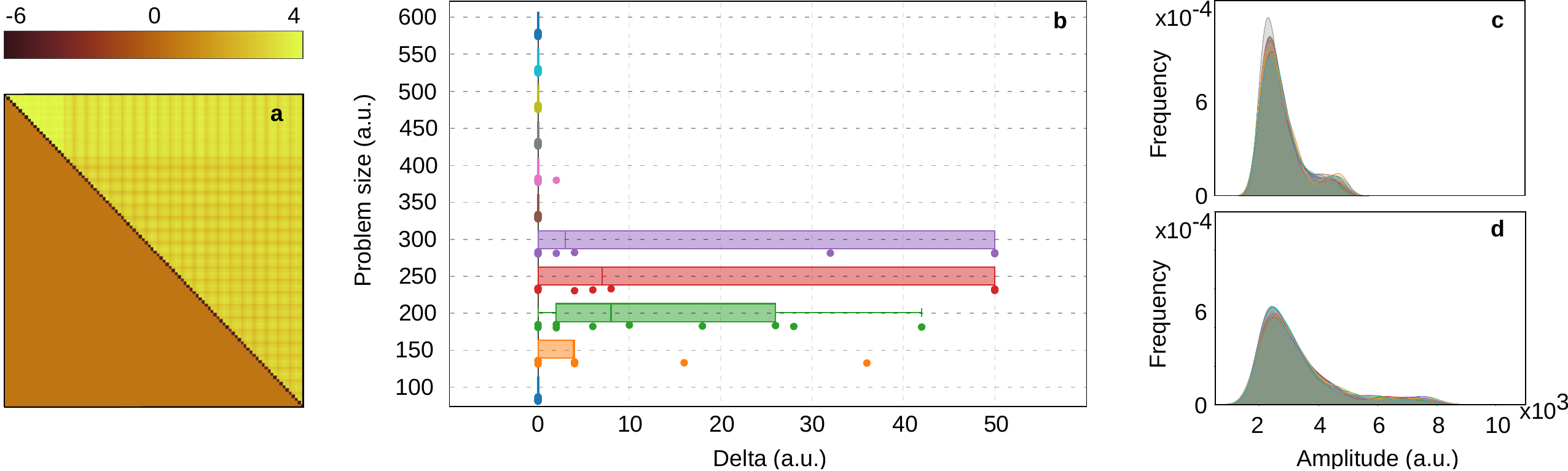}
	\caption{\textbf{QUBO matrix and \textit{delta} distributions over multiple datasets}. 
		$\textbf{a.}$ QUBO matrix of one instance of data with problem size equal to 100. The entries are scaled and the intensity of colors is used as a means to summarize the main characteristics of the plot: the diagonal is made up of negative values, the lower triangular part is zero and the upper one has no null entries. $\textbf{b.}$ Boxplots of deltas for different input problem sizes computed over $10$ datasets for each size with dots representing the values of the delta in each instance. $\textbf{c-d.}$ Kernel density estimation of the distribution of input data for problems with, respectively, $200$ and $500$ variables, showing the data from all $10$ instances in each plot. For \textbf{b}, \textbf{c} and \textbf{d} the values of deltas were saturated to a reference value of $50$, therefore such numerical value is to be interpreted as the result of a bad solution.}
	\label{fig:2}
\end{figure*}

Fig.~\ref{fig:2}a shows the QUBO matrix defining the connectivity of qubits required by the specific NPP instance and with regular patterns related the number amplitudes in the dataset. With the problem being formulated as an Ising model, all variables are coupled in pairs, resulting in a dense (upper-triangular) QUBO matrix. Such connectivity is the most complex to handle and can thus be an issue for current quantum hardwares, making it interesting to investigate the quantum annealer performance.

The distribution of partition deltas for each different problem size is summarized in Fig.~\ref{fig:2}b. We produced $10$ different datasets to be partitioned for every problem size and we computed the value of delta for all these instances. For each problem size we have built a boxplot of deltas centered on the median of the $10$ delta values coming from the solution of the NPP.

The combination of quantum annealing with the classical minor-embedding heuristics is able to find the optimal solution in most cases. This is achieved especially when the problem is very small (and, as a consequence, computationally easy) or when its size is significantly higher. In fact, for our smallest problem and for those with input size greater than $450$ binary variables, we are able to optimally solve the $10$ different NPP instances. On the other hand, for middle-sized problems, not all distributions of data allow qubits to reach the ground state. As a result, we obtain the optimal solutions only for a subset of the given problems.

Figs.~\ref{fig:2}c-d report the density distribution of each of the 10 datasets used for two different problem sizes ($200$ and $500$ variables). As explained above, the quality of the results on the bigger model exceeds the one on intermediate sizes. Comparing both density distributions we can conclude that this behavior is fundamentally related to the fact that a shift of the distribution curve to lower values leads to a dataset containing more solution degeneracy for lower energy states and consequently simpler to solve even when the problem is bigger.  

An effective method to enhance the exploration of the solution space is the direct manipulation of
the annealing schedule \cite{ottaviani2018low, PhysRevApplied.11.044083}.
This distinctive technique can be used to improve the quality of the solution in the cases described before in which we could not reach the ground state.
Indeed, in contrast to what we did with the first approach, where the annealing has been used without interfering with the spontaneous process, we exploit now the capability of the D-Wave solver API to manipulate directly the scheduling of the cycle. To accomplish this, we define the time instant at which the cycle has to be stopped and resumed, as well as the value of the persistent current powering the adiabatic relaxation. This entire procedure is referred to as annealing pause.

\begin{figure}[H]
	\centering
	\includegraphics[width=0.48\textwidth]{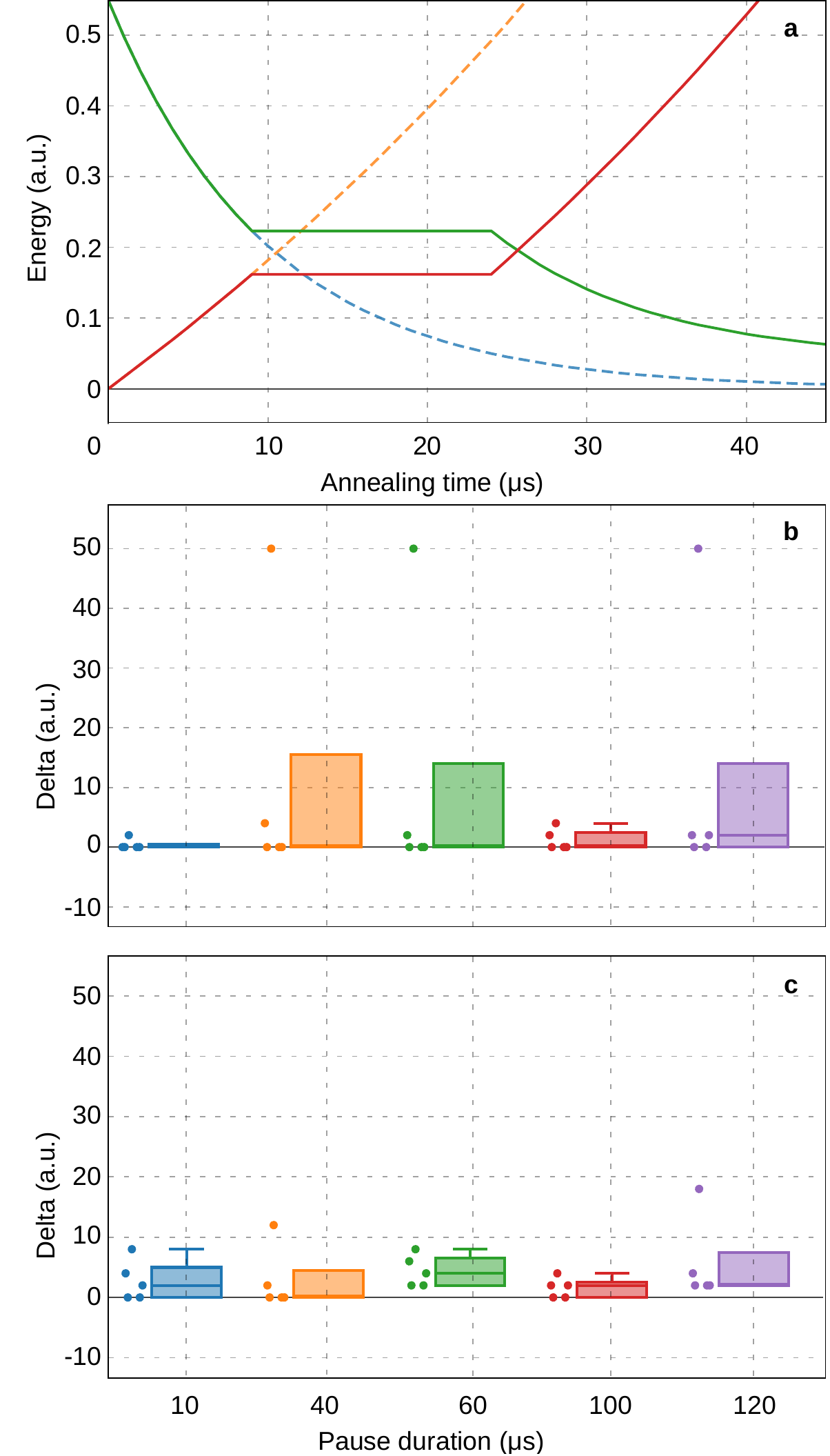}
	\caption{\textbf{Annealing cycle and boxplots of deltas with pause}. 
		$\textbf{a.}$ Sketch representation of the terms contributing to the system Hamiltonian as a function of the time. The solid red, dashed orange, solid green and dashed blue lines represent respectively the problem Hamiltonian with and without pause and the initial Hamiltonian with and without pause. $\textbf{b-c.}$ Boxplots of deltas found over multiple runs of the annealing with the same pause starting point, same value of persistent current but different pause duration times.
		The dots represent the values of deltas for each run (saturated where needed as in Fig.~\ref{fig:2}). \textbf{b.} shows the boxplots for a single instance of the problems  ($300$ variables) associated with 
		a degradated solution, i.e. far from the ground state, whereas \textbf{c.} depicts the same but for a single instance of the problem with $200$ variables.
	}
	\label{fig:3}
\end{figure}

The top panel of Fig.~\ref{fig:3} is a sketch of the evolution over time of the initial and problem Hamiltonians, as the time-scheduled moves forward, compared with their theoretical state if no pause is scheduled. In both cases the problem Hamiltonian grows while the initial one decreases, but in the time-schedule case we find a moment (determined by the user), when the annealing is paused and, as a consequence, the two terms of the system Hamiltonian remain constant. Once the pause is finished, the normal scheduling is resumed and continues its cycle. At the end of the process, the initial Hamiltonian vanishes and the energy of the system is determined by the problem alone.

The middle and bottom panels of Fig.~\ref{fig:3} show the results of the analysis on two problems, one of size 200 and the other of size $300$, for which the uncontrolled annealing performed worst. For each problem we have paused the annealing after $10$ $\mu s$, let the system rest for $10$, $40$, $60$, $100$ and $120$ $\mu s$ respectively halfway through the flow of current and finally let the annealing end. This whole process was repeated five times for each problem.

The best energy configuration in terms of distance from the ground state for the two problems analysed here were not achieved with the same parameters settings.
In fact, every instance requires different values of the pause starting point, duration and persistent current.
Nevertheless, all of our choices greatly improved the results previously obtained with the uncontrolled annealing, even though not all of them led to the optimal solution.
We were able to record considerable results multiple times, proving that the introduction of the pause can increase the accuracy of the annealing.

This improvement in the quality of the results is due to the effect of the pause on the search region of the solution space: by pausing the flow of the persistent current, and hence the annealing, we are able to widen the exploitation of the energy landscape and, as a consequence, the probability of finding the global minimum.

The parameters must be tuned wisely: too long pauses could make the system escape from energy points near the ground state, while too short pauses could not be effective at all. At the same time, if we schedule a pause after the system overcome the minimum gap between the energy levels, i.e: it has already been in some equilibrium state, we will get no benefit from this whole procedure; likewise, a pause scheduled too early will have no effect on the probability of obtaining the global minimum because the chances of escaping the ground state are still high.

It has been shown empirically that finding the appropriate time to start the pause and its duration is a technique that is likely to increase the computation performance of the quantum annealing, yielding much better solutions at the cost of only little more QPU computational time \cite{ottaviani2018low}.

\section{Conclusions}

In this work we studied the capabilities of the D-Wave quantum annealer and the D-Wave Hybrid framework to approach problems in a complex scenarios.
For our analysis we selected a fully connected model: the NPP, which poses an enormous challenge to the currently available QPUs and the architecture they are based on. 
 
Two different analyses were done: accuracy of the outcome when the input size scales up and the impact of the annealing pause on the solution quality.

For the first part we conducted our analysis on a number of small-to-large size problems to investigate the behaviour of the quantum annealer on a level of complexity which is potentially that of real-life problems. One interesting result was found: a discontinuous accuracy with the problem size. While high-quality results were found at small problems, there is a counter-intuitive behaviour as the problem dimension increases: a dip in the accuracy for medium-sized problems and a recovery as it continuous increasing. This effect was explained by the value distribution within the dataset: lower values in the input allow higher accuracy of the result, even when the size of the problem is rising.

The medium-sized problems were studied in more detail by applying pauses during the annealing cycle, allowing the system to explore the solution space with a modified equilibrium. 
Our results prove that with the correct parameters tuning it is possible to improve dramatically the accuracy of the solution, obtaining optimal results in cases that had proven to be troublesome in a 
non-altered context.

\section*{Acknowledgments}
\begin{acknowledgments}
	We acknowledge the support of the Universities Space Research Association, Quantum AI Lab Research Opportunity Program. Also, we thank Davide Venturelli for fruitful discussions.
\end{acknowledgments}

%%% Bibliography %%%%
%%%%%%%%%%%%%
\bibliography{rprj}{}
\bibliographystyle{apsrev}

\balancecolsandclearpage

\end{document}